\documentstyle[12pt]{article}

\newcommand{\newsection}{    
\setcounter{equation}{0}
\section}
\newcommand{\tr}[1]{\,{\rm tr}\,#1\,}
\def\e{{\,\rm e}\,}

\def\eop{\vspace*{\fill}\pagebreak}
\def\be{\begin{equation}}
\def\ee{\end{equation}}
\def\bea{\begin{eqnarray}}
\def\eea{\end{eqnarray}}

\newcommand{\rf}[1]{(\ref{#1})}
\newcommand{\eq}[1]{Eq.~(\ref{#1})}

\def\l{\lambda}
\def\h{\eta}
\def\m{m_0^2}

\newcommand{\ie}{{\it i.e.}\ }

\newcommand{\half}{{\textstyle{1\over 2}}}
\renewcommand{\d}{{{\partial}}}

\newcommand{\ra}{\rightarrow}
\newcommand{\fr}[2]{{\textstyle {#1 \over #2}}}

\begin{document}

\begin{flushright}
ITEP-YM-7-92 \\ August, 1992
\end{flushright}
\vspace{0.5cm}

\begin{center}
{\LARGE Adjoint Fermions Induce QCD} \end{center} \vspace{.5cm}
\begin{center}
{\large S.\ Khokhlachev} \\
\mbox{} \\ {\it Cybernetics Council, Academy of Science} \\
{\it Vavilov st. 40,  117333 Moscow, Russia} \\
\vspace{0.5cm} \mbox{} \\ {\large and} \\
\vspace{0.5cm} \mbox{} \\
{\large Yu.\ Makeenko}\footnote{E--mail: \ makeenko@vxitep.itep.msk.su \ /
\ makeenko@desyvax.bitnet \ / \ makeenko@nbivax.nbi.dk } \\ \mbox{} \\
{\it Institute of Theoretical and Experimental Physics} \\
{\it B.Cheremuskinskaya 25, 117259 Moscow, Russia}
\end{center}

\vspace{1cm}

\begin{abstract}

We propose to induce QCD by fermions in the adjoint representation
of the gauge group $SU(N_c)$ on the lattice. We
consider various types of lattice fermions: chiral, Kogut--Susskind and
Wilson ones. Using the mean field method we show that a first order
large-$N$ phase transition occurs with decreasing fermion mass.
We conclude, therefore, that adjoint fermions induce QCD. We draw the same
conclusion for the adjoint scalar or fermion models at large
number of flavors $N_f$ when they induce a single-plaquette lattice gauge
theory. We find an exact strong coupling solution for the adjoint fermion
model and show it is quite similar to that for the Kazakov--Migdal model with
the quadratic potential.  We discuss the possibility for the adjoint
fermion model to be solvable as $N_c\ra\infty$ in the weak coupling region
where the Wilson loops obey normal area law. \end{abstract}


\eop

\newsection {Introduction}

This paper is inspired by a recent work by Kazakov and Migdal~\cite{KM92}
who proposed to induce QCD by a scalar field on the
lattice.  This scalar field is taken in the adjoint representation of the
gauge group $SU(N_c)$ while the gauge field is attached in the usual way to
make the model gauge invariant except no kinetic term for the gauge field.
This construction makes the model solvable in the large-$N_c$ limit.  The
latter circumstance differs the Kazakov--Migdal model from the previous
attempts to induce Wilson lattice gauge theory by fundamental-representation
scalars~\cite{Ban83} or fermions~\cite{Ham83} at the number of flavors (or
different species) $N_F\sim N_c$.

A price for the solubility is an extra local $Z_N$ symmetry which has to be
spontaneously broken~\cite{KSW92} under the way to the continuum limit.  As
we proposed in Ref.~\cite{KhM92}, if the large-$N$ phase transition occurs
for the Kazakov--Migdal model before the one separating the confinement and
perturbative Higgs phases, one gets normal confinement (area law) in the
intermediate region bounded by the two phase transitions. The problem is,
however, that this scenario in not realized for the original model of
Ref.~\cite{KM92}.

The mean field analysis~\cite{KhM92} indicates that no first order phase
transition occurs, with decreasing mass parameter, before the gaussian model
becomes unstable which is compatible with the
exact solution of Refs.~\cite{Mig92a,Gro92} obtained
solving the master field equation and reproduced by means of loop
equations~\cite{Mak92}. The instability is due to an
unlimited condensation of scalar particles for an action which is unbounded
from below and were lead to the Higgs phenomenon if a stabilizing
self-interaction would be added.

The latter fact prompts the following way out. Let us consider
fermions rather than bosons. On the one hand, the induced lattice
action obtained by the large mass expansion looks for fermions very similar
to that for scalars. On the other hand, since fermions never condense, the
gaussian fermionic action in an external field is stable so one can
arbitrarily decrease the mass parameter. Moreover, if one assumes that the
gauge field is weakly fluctuating, the perturbative expansion for fermions
looks like that for QCD and we expect, therefore, a phase
transition which separates the perturbative region from the strong coupling
phase with unbroken $Z_N$ symmetry and associated local confinement.  This
phase transition is to be identified with the large-$N$ phase transition
discussed above.

In the present paper we propose to induce QCD {\it a la} Kazakov--Migdal by
fermions in the adjoint representation of the gauge group on the lattice. We
consider various types of lattice fermions: chiral, Kogut--Susskind and
Wilson ones. We discuss first
the adjoint scalar and fermionic models at large number of
flavors $N_f$ and show they induce a single-plaquette lattice gauge theory
quite similarly to Refs.~\cite{Ban83,Ham83} for the fundamental
representation.
Using the mean field method we show that a first order
large-$N$ phase transition occurs for chiral and Kogut-Susskind fermions
at $N_f=1$. We conclude, therefore, these two models induce QCD.
The similar conclusion presumably holds
for Wilson fermions while the results for $N_f=1$ are less certain.
We find an exact strong coupling solution of loop equations for the
adjoint fermion model and show that for chiral fermions it is quite similar
to the solution~\cite{Gro92} for the Kazakov--Migdal model with the
quadratic potential while for Wilson fermions the exact result coincides
with the leading order of the large mass expansion.  We point out that, as
usual, the strong coupling solution is not sensitive to the first order
large-$N$ phase transition which occurs {\it inside\/} the region where the
large mass expansion is convergent.  We discuss the possibility for the
proposed models to be solvable as $N\ra\infty$ in the weak coupling region
where the Wilson loop averages obey normal area law.

\newsection{Inducing single-plaquette action at large $N_f$}

The scenario of breaking the $Z_N$ discussed in Section~1 can be completely
realized at large number of flavors $N_f$ when the induced gauge theory
is in our case the single-plaquette lattice gauge theory with the adjoint
action quite similarly to Refs.~\cite{Ban83,Ham83} for the
fundamental representation of $SU(N_c)$. The large-$N$ phase transition
occurs for the former action according to the pattern we proposed long time
ago~\cite{KhM81}.

\subsection{Scalars}

For simplicity let us begin with the case of scalars.
The extension of the Kazakov--Migdal model~\cite{KM92} to the case of $N_f$
flavors reads
\be
Z=\int \prod_{x,\mu} dU_{\mu}(x) \prod_x \prod_{f=1}^{N_f} d\Phi_f(x)
\e^{\sum_{f=1}^{N_f} \sum_x N_c \tr{\left(-m_0^2\Phi_f^2(x)+
\sum_{\mu}\Phi_f(x)U_\mu(x)\Phi_f(x+\mu)U_\mu^\dagger(x)\right)}}
\label{spartition}
\ee
where $N_f$ fields $\Phi_i(x)$ ($i=1,\ldots,N_f$) take values in the
adjoint representation of the gauge group $SU(N_c)$ and the link variable
$U_\mu(x)$ belongs to the gauge group.  For the purposes of the present paper,
we
restrict ourselves by the quadratic potential. The original Kazakov--Migdal
model corresponds to $N_f=1$.

Calculating $N_f$ gaussian integrals, one rewrites the model~\rf{spartition}
as the lattice gauge theory
\be
Z=\int \prod_{x,\mu} dU_\mu(x)\,\e^{-S_{ind}[U_\mu(x)]},
\label{gaugetheory}
\ee
with the induced action given by the following large mass expansion
\be
S_{ind}[U]=-\frac {N_f}{2} \sum_{\Gamma}
\frac{|\tr{U(\Gamma)}|^2}{l(\Gamma)m_0^{2l(\Gamma)}}.
\label{ind}
\ee
For $N_f=1$ one recovers the result of Ref.~\cite{KM92}.

We apply now the idea by Bander~\cite{Ban83} of how to obtain the
single-plaquette lattice action from the sum over loops on the r.h.s.\ of
\eq{ind}. Let $N_f$ amd $m_0^2$ simultaneously tend to $\infty$ so that
$m_0^2\sim (N_f)^{\fr 14}$. Then the only single-plaquette term survives
while those associated with longer loops are suppressed at least as
$(N_f)^{-\fr 12}$. Therefore we have shown in this limit that
\be
S_{ind}= -\frac{N_f}{8m_0^8} \sum_p \left|\tr{U(\d p)}\right|^2
\label{adjaction}
\ee
where $\d p$ is the boundary of the plaquette $p$. In other words the
induced theory is just the single-plaquette lattice gauge theory with the
adjoint action.

As is known~\cite{KhM81} the lattice gauge theory with the
action~\rf{adjaction} undergoes the (first order) large-$N$ phase
transition at $N_f/(4m_0^8)\approx 2$. This is exactly the phase
transition advocated for the Kazakov--Migdal model by Kogan, Semenoff and
Weiss~\cite{KSW92} and needed to induce QCD according to the scenario of
Ref.~\cite{KhM92}.  Below the phase transition, the adjoint
action~\rf{adjaction} is reduced~\cite{KhM81} at $N_c=\infty$ to a
fundamental single-plaquette action and the Wilson loop averages obey area law.

It is crucial for this consideration that the critical value
\be
m_*^2\approx\left( \frac{N_f}{8} \right)^{\fr 14} >D
\label{m_*}
\ee
as $N_f\ra\infty$ so that the gaussian model \rf{spartition} is stable.
A mechanism to escape the instability while having the large-$N$ phase
transition is pretty simple: the contribution of long loops in the
effective action \rf{ind} which were lead to condensation is now suppressed
by construction. This pattern looks like the idea to avoid
instability by means of fermions since in the fermionic case the long loops
do not condense due to alternating signs in the sum (see \eq{find} below).

On the contrary there is no such phase transition for $\m>D$ at $N_f=1$.
This statement relies on the mean field study of Ref.~\cite{KhM92}
and is compatible with
the exact solutions~\cite{Gro92,Mak92}. Therefore, we expect the
large-$N$ phase transition to occur for the model~\rf{spartition} only for
$N_f$ larger than some value $N_{f*}$. This value is estimated in Section~3.5
by the use of the mean field method.

\subsection{Fermions}

Similarly to how Hamber~\cite{Ham83} used large-$N_f$ limit to induce the
Wilson single-plaquette lattice gauge theory by fermions in the fundamental
representation of $SU(N_c)$, we can extend the results of the previous
section to the fermionic case.

Let us consider the {\it adjoint fermion model} (AFM) which is defined by
the partition function
\be
Z_{AFM}=\int \prod_{x,\mu} dU_{\mu}(x) \prod_x \prod_{f=1}^{N_f} d\Psi_f(x)
d\bar{\Psi}_f(x) \e^{-S_{F}[\Psi,\bar{\Psi},U]}.
\label{fpartition}
\ee
Here $S_{F}[\Psi,\bar{\Psi},U]$ is the
standard action for fermions on the lattice:
\bea
 & & S_{F}[\Psi,\bar{\Psi},U] =
\sum_{f=1}^{N_f} \sum_x N_c \tr{\left(m\bar{\Psi}_f(x)\Psi_f(x) \right.
\nonumber \\* & & \left.-
\sum_{\mu=1}^D [\bar{\Psi}_f(x)P_\mu^-U_\mu(x)\Psi_f(x+\mu)U_\mu^\dagger(x)}
+\bar{\Psi}_f(x+\mu)P_\mu^+U_\mu^\dagger(x)\Psi_f(x)U_\mu(x)]\right)
\label{faction}
\eea
where $\Psi_f(x)$ is the Grassmann anticommuting field while
\be
P_\mu^\pm=r\pm\gamma_\mu
\label{projectors}
\ee
stand for the projectors. The case $r=0$ corresponds to chiral fermions
while $r=1$ is associated with Wilson fermions. As is well known, the chiral
fermions describe $2^D N_f$ flavors in the naive continuum limit while
Wilson fermions are associated with $N_f$ flavors.

The path integral over $\Psi$ can be calculated by the large mass expansion
similarly to the scalar case. One gets for the induced action
\be
S_{ind}[U] = - N_f \sum_{\Gamma}
\frac{|\tr{U(\Gamma)}|^2}{l(\Gamma)m^{l(\Gamma)}}\, \hbox{Sp}\,
\prod_{l\in \Gamma} P_\mu
\label{find}
\ee
where \ Sp \ stands for the trace over the spinor indices of the
path-ordered product of the projectors $P_\mu^\pm$ (plus or minus depends on
the orientation of the link $l$) along the loop $\Gamma$.

The contribution of leading order in $1/m$ comes from a plaquette which gives
\be
S_{ind}= -\frac{2^{\frac D2}N_f(1+2r^2-r^4)}{m^4}
\sum_p \left|\tr{U(\d p)}\right|^2.
\label{fadjaction}
\ee
If $m\sim N_f^{\fr 14}$, only this term survives in the large-$N_f$ limit
while the others are suppressed at least as $N_f^{-\half}$ similarly to the
scalar case.

The estimate of the critical values of $m$ which are associated with the
large-$N$ phase transition gives $m_*\approx(2N_f)^{\fr 14}$ for chiral
fermions and $m_*\approx(4N_f)^{\fr 14}$ for Wilson fermions. As we shall
see in the next section, the mean field analysis indicates that the phase
transition certainly exists in the former case at $N_f=1$ as well.

\newsection{Mean field analysis}

While we have shown in the previous section that both adjoint scalar and
adjoint fermion models induce the single-plaquette Wilson lattice gauge
theory in the limit of large $N_f$, there is no hope for this limit to be
solvable even at $N_c=\infty$ (this were correspond to an exact solution of
Wilson lattice gauge theory at $N_c=\infty$).
We investigate in this section AFM with $N_f=1$ which, as is discussed in
Section~5, could be as solvable as the Kazakov--Migdal model.

\subsection{Mean field with fermions}

We apply to AFM defined by the partition function \rf{fpartition} at $N_f=1$
the (variational) mean field approach of Ref.~\cite{KhM92}. Since we
substitute the mean field solely for the gauge field and exactly integrate
over fermions, we expect the mean field method is applicable for our case
despite the fermionic system.

Substituting the mean field value $[U_\mu(x)]_{ij}=\h\delta_{ij}$ for all
links of the lattice except the given one, we get the following
self-consistency condition
\be 2(\h-\h^2)=\frac{1}{b_A}
\label{wc} \ee where
\be b_A= \frac{\int \prod_{x}d\Psi(x)d\bar{\Psi}(x) \e^{
-S_{F}[\Psi,\bar{\Psi},\h]} \fr 1N
\tr{\left(\bar{\Psi}(0)P_\mu^-\Psi(0+\mu)
+\bar{\Psi}(0+\mu)P_\mu^+\Psi(0)\right)}}
{\int \prod_{x}d\Psi(x)d\bar{\Psi}(x)
\e^{-S_{F}[\Psi,\bar{\Psi},\h]}}
\label{b_A}
\ee
and
\bea
 & & S_{F}[\Psi,\bar{\Psi},\h] = \sum_x N\tr{\left(m\bar{\Psi}(x)\Psi(x)
\right.\nonumber \\ & & \left.
-\h^2\sum_{\mu=1}^D [\bar{\Psi}(x)P_\mu^-\Psi(x+\mu)}
 + \bar{\Psi}(x+\mu)P_\mu^+\Psi(x)]\right) .
\eea
\eq{wc} holds providing $\h>1/2$, which is prescribed by the requirement for
the corresponding one-matrix model to possess a weak coupling solution.
To find the location of the phase transition one uses either {\it geometric}
criterion which is based on terminating the $\h\neq0$ solution with
increasing $m$ or the {\it thermodynamic} criterion which says that the
phase transition occurs when free energies of the $\h=0$ and $\h\neq0$ phases
coincide.

\subsection{Chiral fermions}

For chiral fermions the gaussian integral on the r.h.s.\ of \eq{b_A} can
easily be calculated by the standard lattice technique (see
Ref.~\cite{Wil75}) to give
\be
b_A=\frac{2^D}{\h^2}\prod_{\mu=1}^D \int_{-\pi}^{\pi} \frac{dp_\mu}{2\pi}
\frac{1-\fr 1D \sum_\mu\cos{p_m}}{\fr{m^2}{2\h^4}+D-\sum_\mu\cos{p_\mu}}=
\frac{2^D}{D\h^2}\left(1-\frac{m^2}{2\h^4}\int_0^\infty d\alpha
\e^{-\left(\fr{m^2}{2\h^4}+D\right)\alpha}\,\hbox{I}_0^D(\alpha)\right).
\label{bchiral}
\ee
Here \ I$_0(\alpha)$ is the modified Bessel function.

The r.h.s.\ of \eq{bchiral} multiplied by $\h^2$
monotonically increases with decreasing $m$
approaching the maximal value $2^D/D$ at $m=0$. Moreover, it can be
approximated with great accuracy by the first term of the expansion in
$x=1/(\fr {m^2}{2\h^4}+D)$ which can be seen from the series expansion
\bea
\h^2 b_A=16x-8x^2+32x^3-42x^4+168x^5-320x^6+1280x^7-\fr{23765}{8}x^8
+\fr{23765}{2}x^9 \nonumber \\
-\fr{124047}{4}x^{10}+124047x^{11}-\fr{1397319}{4}x^{12}+
1397319x^{13}+{\cal O}(x^{14}).
\eea
Thus, we substitute
\be
\frac{1}{b_A} \approx 2^{-D}(\fr {m^2}{2\h^4}+D)\h^2
\label{approximation}
\ee
and there is no need in numerical calculations.

The critical value $m_*$ can be easily determined from the above
mentioned geometric criterion which says
\be
\left.\frac{\d \h}{\d m}\right|_*=\infty \hbox{ \ \ \ \ or \ \ \ \ }
\left.\frac{\d m}{\d \h}\right|_*=0
\label{geometric}
\ee
at the critical point $m_*$. Differentiating \eq{wc} with the r.h.s.\
substituted by \eq{approximation}, one gets the solution
\be
m_*\approx 6.5, \hbox{ \ \ \ and \ \ } \h_*\approx \frac 23
\label{results}
\ee
at $D=4$.

Some comments about the
obtained critical values of the large-$N$ phase transitions in the case of
chiral fermions are in order.
\begin{itemize}
\vspace{-7pt} \item
In contrast to the scalar case analyzed in Ref.~\cite{KhM92}, the integral
on the r.h.s.\ of \eq{bchiral} never diverges for $0<m^2<\infty$ which is a
consequence of the stability of the fermionic system discussed above.
Since $\h^2 b_A$ can now reach values larger than $0.5$, \eq{wc} always
possesses a
solution.
\vspace{-7pt} \item
The coefficient $2^D$ on the r.h.s.\ of \eq{bchiral} is nothing but the
number of flavors which is associated with the naive continuum limit of
chiral lattice fermions. Therefore, one could relate the appearance of the
large-$N$ phase transition with an effective number of flavors similarly to
Section~2.2. This point of view explains why, as is shown below, the
large-$N$ phase transition still certainly occurs for Kogut--Susskind fermions
describing $2^{\fr D2}$ flavors and presumably for Wilson lattice
fermions which correspond to $1$ flavor in the naive continuum limit.
 \vspace{-7pt} \end{itemize}

\subsection{Kogut--Susskind fermions}

As is well known, the number of continuum flavors described by chiral
fermions can be reduced by $2^{\frac D2}$ passing to Kogut--Susskind
fermions. To this aim one performs the following transformation of the
spinor $\Phi(x)$
\be
\Psi(x)=(\gamma_1)^{x_1}(\gamma_2)^{x_2}\ldots(\gamma_D)^{x_D}
\chi(x)
\ee
where the coordinates of the lattice vector $x_\mu$ are measured in the
lattice units. In the new variables the fermionic lattice action \rf{faction}
reads
\bea
& & S_{F}[\chi,\bar{\chi},U] =
 \sum_x N_c \tr{\Big(m\bar{\chi}(x)\chi(x) \nonumber \\* & &
-\sum_{\mu=1}^D\eta_\mu(x)[\bar{\chi}(x)U_\mu(x)\chi(x+\mu)U_\mu^\dagger(x)}
 -\bar{\chi}(x+\mu)U_\mu^\dagger(x)\chi(x)U_\mu(x)]\Big)
\label{ksaction}
\eea
where
\be
\eta_1(x)=1; \;\;\;     \eta_\mu(x)=(-1)^{x_1+\ldots+x_{\mu-1}}.
\label{eta}
\ee
Since the action \rf{ksaction} is diagonal w.r.t.\ spinor indices, one can
keep only one component of $\chi(x)$ so that it can be considered as a scalar
Grassmann variable. The normal spinors are reproduced in the continuum limit
by $2^D$ $\chi$'s taken at neighbor sites of the lattice. Such a procedure
manifestly reduces the number of flavors in the continuum by $2^{\fr D2}$.

The mean field analysis of AFM with Kogut--Susskind fermions is quite
similar to that of the previous section. One should only replace $2^D$ in
Eqs. \rf{bchiral} end \rf{approximation} by $2^{\fr D2}$ so that the
self-consistency condition~\rf{wc} at $D=4$ reads
\be
2\left(\frac 1\h -1 \right) =\frac{m^2}{8\h^4} +1 .
\ee
This equation determines the critical values associated with the large-$N$
phase transition to be
\be
m_*\approx 0.7, \hbox{ \ \ \ and \ \ } \h_*\approx \frac 12.
\label{ksresults}
\ee

Therefore, we have shown that the large-$N$ phase transition occurs for AFM
with Kogut--Susskind fermions. The critical values~\rf{ksresults} are
smaller than those given by~\eq{results} and are near the limit of \sloppy
applicability.

\subsection{Wilson fermions}

For AFM with Wilson fermions $b_A$ is given by \eq{b_A}. The standard
calculation~\cite{Wil75} of the gaussian fermionic integral yields
\be
b_A=\frac{2^\fr D2}{D\h^2}\prod_{\mu=1}^D \int_{-\pi}^{\pi}
\frac{dp_\mu}{2\pi}
\left(1- \frac{m}{2\h^2}
\frac{\fr{m}{2\h^2} -\sum_{\mu}\cos{p_\mu}}
{\left( \fr{m}{2\h^2} -\sum_{\mu}\cos{p_\mu}\right)^2+\sum_\mu
\sin^2{p_\mu}}\right) .
\label{bwilson}
\ee

The integral on the r.h.s.\ of \eq{bwilson} possesses a number of
interesting properties. Its large mass expansion
\be
\h^2 b_A=6\left(\frac{2\h^2}{m}\right)^4 + 66\left(\frac{2\h^2}{m}\right)^6
+{\cal O}\left(\left(\frac{2\h^2}{m}\right)^8\right)
\ee
begins with the term $\sim m^{-4}$ rather than $m^{-2}$ as for chiral
fermions or bosons. The point is that {\it backtracking} paths in the sum
over paths associated with $b_A$ are now forbidden since the
product of projectors
\be
P_\mu^+P_\mu^-=r^2-1
\label{zero}
\ee
vanishes for $r=1$. For this reason our mean field coincides for Wilson
fermions with the one which would be obtained directly from the induced
action~\rf{find} where the backtracking paths never contribute due to
unitarity of $U$. A price for this nice property is the complicated integral
on the r.h.s.\ of \eq{bwilson}.

Since $b_A$ varies between $0$ at $m=\infty$ and $1$ at $m=0$, one expects
Eqs.  \rf{wc} and \rf{geometric} to possess a solution for the critical
value $m_*$. We performed an approximate numerical analysis and found
$m_*\approx \h_* \approx 0.5$ which looks like the case of Kogut--Susskind
fermions. However, now an uncertainty of the results is bigger so that our
conclusion is that, while the large-$N$ phase transition presumably occurs
for Wilson fermions, more accurate investigation maybe by other methods is
needed. However, the large-$N$ phase transition certainly occurs for AFM
with Wilson fermions for $N_f\geq 2$.

\subsection{$N_f$ scalars}

The mean field method can be applied to estimate
the critical value $N_{f*}$ below which the large-$N$ phase transition
discussed in Section~2.1 disappears.  The corresponding self-consistency
condition is given by \eq{wc} where $b_A$ reads
\be
b_A=\frac{N_f}{2\h^2}\int_0^\infty d\alpha
\,\e^{-\frac{m_0^2}{\h^2}\alpha} \; \hbox{I\/}_0^{D-1}(\alpha)
\; \hbox{I\/}_1(\alpha)
\label{tildeb}
\ee
and \/I$(\alpha)$ being the modified Bessel function.

The condition $\h>1/2$, which is prescribed by the requirement for the
corresponding one-matrix model to possess a weak coupling solution,
yields $\eta^2 b_A>1/2$ or $N_{f*}>30$ --- a pretty big
value --- since the maximal value of the
integral on the r.h.s.\ of \eq{tildeb} is $0.06$ at $D=4$ .

\newsection{Loop averages at strong coupling}

AFM can be exactly solved at $N_c=\infty$ in the strong coupling limit (\ie
before the large-$N$ phase transition). We obtain the result solving loop
equations quite similarly to Ref.~\cite{Mak92} where
the Kazakov--Migdal model with the quadratic potential was solved by this
method.  The solution for AFM with chiral fermions obtained below is quite
similar to that by Gross~\cite{Gro92}
for the Kazakov--Migdal model with the quadratic potential and coincides
with the old results by Klugberg-Stern {\it et al.}~\cite{Chi} for the
infinite-coupling limit of lattice $N_c=\infty$ QCD with chiral fermions.
We show this fact is not a coincidence and is related to very peculiar
properties of Wilson loop averages in AFM model as well as in the
Kazakov--Migdal model. For Wilson fermions the exact solution of loop
equations coincides for the same reason with the leading order of the
large mass expansion. It will be explicitly demonstrated this strong
coupling solutions are {\it not\/} singular, as usual,  at the point of the
first order large-$N$ phase transition which is determined by another
criterion.

\subsection{Loop equations in AFM}

All the observables in AFM can be expressed via the adjoint Wilson
loop
\be
W_A(C)=\left\langle\frac{1}{N_c^2}
\left( \left| \tr{U(C)}\right|^2-1\right)\right\rangle\,,
\label{adjloop}
\ee
where the average is understood with the same measure as in \eq{fpartition},
and the fermionic path-dependent amplitude \be G_{ij}(C_{xy})= \Big\langle
\frac {1}{N_c} \tr{\left( \Psi_i(x)U (C_{xy}) \bar{\Psi}_j(y)
U^\dagger(C_{xy}) \right)} \Big\rangle \label{G} \ee where $i$ and $j$ are
spinor indices.

The amplitude $G_{ij}(C_{xy})$ obeys two sets of loop equations which are
analogous to the loop equations in QCD with quarks (in the fundamental
representation of the gauge group). It is simple to obtain an equation
resulting from the Dirac equation for $\Psi(x)$ which reads \be m G(C_{xy})
-\sum_{\mu=1}^D [ P_\mu^+G(C_{(x+\mu)x}C_{xy})
+P_\mu^-G(C_{(x-\mu)x}C_{xy})]
=\delta_{xy} W_A(C_{xy})
\label{sd}
\ee
where the matrix multiplication over spinor indices is implied.
Here the path $C_{(x+\mu)x}C_{xx}$ is obtained by attaching the link
$(x,\mu)$ to the path $C_{xx}$ at the end point $x$ as is depicted in
Fig.~1. The projectors are given by \eq{projectors} for chiral and Wilson
fermions or by \eq{eta} for Kogut--Susskind fermions. Due to the presence of
the delta-function, $W_A$ on the r.h.s.\ of \eq{sd} always enters for closed
loops, \ie it is an object of the type~\rf{adjloop}.

Similarly to Ref.~\cite{Mak92} where an analog of \eq{sd} was solved for
the Kazakov--Migdal model with the quadratic potential at strong coupling,
\eq{sd} admits the
ansatz
\be
G_{ij}(C_{xy})=G_L \left(\prod_{l\in
C_{xy}}P_\mu^\pm \right)_{ij} \label{spinfactor} \ee
where plus or minus
corresponds to the direction of the link $l$ which belongs to the contour
$C_{xy}$ and $L$ is the {\it algebraic\/} length~\cite{Mig83} (\ie the one
after contracting the backtrackings) of $C_{xy}$.  The spin factor is needed
to cancel the projectors in \eq{sd}.

For such an ansatz, \eq{sd} for AFM is reduced
to the following recurrent relations on $G_L$
\bea m G_{L} - G_{L-1} -
(2D-1)\sigma G_{L+1}& =& 0 \hbox{ \ \ \ for \ \ } L\geq1\;, \nonumber \\
m G_0 - 2D \sigma G_1&=&1 \label{ansatz} \eea
where $\sigma=r^2-1$. These equations are quite similar to the corresponding
equations for the Kazakov--Migdal model with the quadratic
potential which are given~\cite{Mak92} by the same formulas with $\sigma=1$
and $m$ replaced by $2m_0^2$.
The drastic simplification of the loop equation \rf{sd} is due to the
fact~\cite{KhM92} that $W_A(C)$ vanishes in the strong coupling region at
$N=\infty$ except for $C$ with vanishing minimal area $A_{min}(C)$ (\ie
contractable to a point owing to unitarity of $U$'s):  \be
W_A(C)=\delta_{0,A_{min}(C)}+{\cal O}\left({1\over N^2} \right) \;.
\label{1overN} \ee

\eq{ansatz} can be solved introducing the generating function
\be G(\l)=\sum_{L=0}^\infty G_L \l^L. \ee
The solution at $\sigma\neq0$ (the solution for $\sigma=0$ \ie for Wilson
fermions is described in the next section) is expressed via $G_0$ which was
up to now arbitrary.  It can be determined imposing~\cite{Mak92} the
single-pole analytic structure of $G(\l)$ as a function of the spectral
parameter $\l$.  This requirement unambiguously determines $G(\l)$ to be
\be
G(\l)=\frac{2(2D-1)\sigma G_0}{
\left(\sqrt{m^2+4(1-2D)\sigma}-m\right) \l +2(2D-1)\sigma}
\label{final}
\ee
and
\be
G_0=\frac{2D-1}{m(D-1)+D\sqrt{m^2+4(1-2D)\sigma}}\;.
\label{gross}
\ee
The single-pole prescription is justified in the next section.

\subsection{Properties of strong coupling solution}

Let us consider the strong coupling solution of AFM,
obtained in the previous section, from the viewpoint of the path
representation of the amplitude~\rf{G}. At $N_c=\infty$ one gets
\be
G_{ij}(C_{xy}) = \sum_{\Gamma_{yx}}
\frac{1}{m^{l(\Gamma)}}W_A(C_{xy}\Gamma_{yx}) \left(\prod_{l\in
\Gamma_{yx}}P_\mu^\pm \right)_{ij}
\label{paths} \ee
where the path $C_{xy}\Gamma_{yx}$ is closed by construction.
\eq{paths} is nothing but the standard large mass
representation~\cite{Wil75} of the amplitude in lattice gauge theories
with fermions while the Wilson loop is taken in the same representation as
fermions (adjoint in our case). The fact that the only one-loop average
enters \eq{paths} is due to $N_c=\infty$~\cite{Mig83}.

In the strong coupling region where $W_A(C)$ is given by the simple
formula~\rf{1overN}, the r.h.s.\ of \eq{paths} drastically simplifies. The
path $\Gamma_{yx}$ must coincide with $C_{xy}$ passing in the opposite
direction modulo backtrackings%
\footnote{One should not confuse these backtrackings of $\Gamma$ with
possible backtrackings of $C$ discussed above. The latter ones can be always
contracted due to unitarity.} since $W_A$ equals 1 in this case and $0$
otherwise.  Therefore, the problem of calculating the amplitude $G$ by the
sum over paths is reduced to a problem of counting backtrackings of
one-dimensional trees embedded in a $D$-dimensional hypercubic lattice. The
above formula~\rf{final} represents the solution of this problem. Notice
that since the spin structure on the r.h.s.\ of \eq{paths} factorizes, the
coefficient functions $G_L$ are quite similar for chiral fermions and
scalars.

We would like to point out that exactly the same problem emerges in lattice
QCD at infinite coupling where quarks are in the fundamental representation
but the Wilson loops for non-contractable paths vanish since the coupling is
infinite. Its solution obtained for chiral fermions by
Klugberg-Stern {\it et al.}~\cite{Chi} using a different method exactly
coincides
with \eq{gross}.

The result of Ref.~\cite{Chi} was reproduced by Mkrtchyan and one of the
authors~\cite{KhMk} by means of loop equations. While the loop equation
resulting from the Dirac equation for fermions looks like \eq{sd} with $W_A$
on the r.h.s. substituted by the fundamental Wilson loop, the
loop equations resulting from the (lattice) Maxwell equation are simplified
since fermions are now in the fundamental representation. All the
definitions in this case are the same as above with the path ordered
phase factor in the adjoint representation substituted by the one in the
fundamental representation, say the definition of the fundamental amplitude
$G$ reads
\be
G_{ij}^F(C_{xy})= \Big\langle
{\Psi}_i^F(x)U (C_{xy}) \bar{\Psi}_j^F(y) \Big\rangle .
\label{G^F}
\ee
The equation resulting from the variation of $U_\mu(x)$ at the link
$(x,\mu)\in C_{xy}$ emanating from the point $x$ reads%
\footnote{The bilinear in $G$ terms were missing in the original
paper by Weingarten~\cite{Wei79}.}
\be
G^F(C_{xy})=G^F(C_{x(x+\mu)})\gamma_\mu G^F(C_{xy}) - G^F(0)\gamma_\mu
G^F(C_{(x+\mu)y}) \label{sd1}   \ee
for chiral fermions. For the ansatz~\rf{spinfactor}
one gets
\be G_L=G_0 G_{L-1}-\sigma G_1 G_L  \hbox{ \ \ \ \ for \ \ \ } L\geq1 .
\label{ansatz1}
\ee

While the corresponding loop equation for AFM does not coincide, generally
speaking, with \eq{sd1}, for the ansatz~\rf{spinfactor} it should reduce, as
is discussed above, to \eq{ansatz1}. At $L=1$ this equation gives one more
relation between $G_0$ and $G_1$ which together with the $L=0$
equation~\rf{ansatz} result in a quadratic equation for $G_0$ having an
unambiguous solution~\rf{gross} expandable in $1/m$ . Therefore, we have
shown that the above strong coupling solution for AFM with chiral fermions
(or for the Kazakov--Migdal model with the quadratic potential) is unique.
In particular, we have justified the single-pole procedure of
Ref.~\cite{Mak92}.

An analogous exact strong coupling solution of loop equations
for AFM with Wilson fermions can be obtained as well.
Since $\sigma=0$ for $r=1$, Eqs. \rf{ansatz} are further simplified so that
the solution reads \be G_L=\frac{1}{m^L} \label{wfinal} \ee which coincides
with the leading order of the large mass expansion.  The reason behind this
it as follows.  As is discussed in Section~3.4, the problem of backtrackings
is missing for Wilson fermions due to the property~\rf{zero} of projectors
$P_\mu^\pm$.  Therefore, the leading term of the large mass
expansion gives the {\it exact} result for AFM with Wilson fermions.

Some comments about the properties of
the exact strong coupling solutions \rf{final} and \rf{wfinal}
of AFM with chiral and Wilson fermions are now in order:
\begin{itemize} \addtolength{\itemsep}{-7pt} \vspace{-7pt}
\item
The solutions~\rf{final}, \rf{wfinal} are not  singular for $m^2>0$
in contrast to the scalar case. Therefore, the strong coupling solution is
fermion case is stable everywhere.
However, AFM undergoes the first order large-$N$
phase transition after which the strong coupling solution is not applicable.
As usual for first order phase transition, this critical point is quite
regular from the viewpoint of the large mass expansion and is determined by
another method (like the mean field approach of Section~3). After the
large-$N$ phase transition occurs, \eq{1overN} that corresponds to local
confinement is no longer valid and should be replaced by normal area law.
The ansatz~\rf{spinfactor} is no longer applicable in this region and the
solution of loop equations would be quite different.
\item
While for $D=1$ the strong coupling solution~\rf{final} coincides with the
free theory amplitudes, it looks {\it simpler\/} for $D>1$ than the
corresponding free theory results given by formulas which are similar to
those of Section~3.
\item
All the statements of Ref.~\cite{Chi} concerning the breaking of chiral
symmetry for QCD with chiral fermions at infinite coupling are applicable to
our case. In particular, no extra phase transition associated with
spontaneous breaking of chiral symmetry occurs with decreasing mass
and $<\bar{\Psi}\Psi>$ which is determined by \eq{gross} is non-vanishing
at $m=0$.
 \item Loop equations turned out to be perfect to solve
the backtracking problem arising in the large mass expansion and
unambiguously determine the solution without any assumptions.  \end{itemize}
\vspace{-6pt}

\newsection{Conclusions and discussion}

The main conclusion we draw from the results obtained in the present paper
is that a simple replacement of scalars by fermions in the Kazakov--Migdal
model cures the problem of instability and provides the large-$N$ phase
transition when the fermion mass is decreased. While in the strong coupling
phase (before the phase transition) one gets local confinement which is
characterized by \eq{1overN}, normal area law holds in the weak coupling
phase (after the phase transition). This phase resembles very much the
standard lattice gauge theory. Moreover, at large-$N_f$ a single plaquette
lattice action arises. Hence, we conclude that AFM induces QCD.

The obtained exact strong coupling solution of AFM does
not know anything about the large-$N$ phase transitions which is associated
with freezing of the gauge field $U_\mu(x)$ near some mean field value $\h$
while $\h=0$ in the strong coupling region. This is usual for first
order phase transitions. The weak coupling solution of AFM should be quite
different from the strong coupling one.

After a nonvanishing value of $\h$ emerges, the arguments of
Ref.~\cite{KM92} about the continuum limit caused by long loops become
applicable. We expect therefore that AFM will induce continuum QCD as well.
One more argument in favor of this is the mentioned above similarity
between the weak
coupling phases of AFM and of the Wilson lattice gauge theory.

A very interesting question is whether the solubility of the Kazakov--Migdal
model by means of the master field equation~\cite{Mig92a} would be pursued
for AFM. This were give a way to solve AFM in the weak coupling region which
would be an alternative to loop equations. As is mentioned already, we do not
think this would be the case for any $N_f$. We see however some hope to
derive an extension of the Itzykson--Zuber--Metha integral, which
played a crucial role in the approach of Refs.\cite{KM92,Mig92a}, to the
 case of Kogut--Susskind like fermions:
 \be
 I[\psi,\chi]=\int dU \e^{N_c\tr{\psi U \chi U^\dagger}}
 \label{FIZ}
 \ee
 where $\psi$ and $\chi$ are Grassmann variables which transform under the
 adjoint representation of $SU(N_c)$. While it is not clear of how to reduce
 $\psi$ and $\chi$ to a diagonal form, the integral over the Haar measure on
 the r.h.s.\ of \eq{FIZ} could be tractable by other methods.
 This problem deserves future investigations.

\section*{Note added}

When this paper was being prepared for publication, there appeared two more
papers \mbox{\cite{new,Mig92d}} on the Kazakov--Migdal model. While
Ref.~\cite{new} deals mostly with the quadratic potential, a strong coupling
solution for a general model with fundamental Wislon fermions is
obtained by Migdal~\cite{Mig92d}.

\eop

\eop

\section*{Figures}
\vspace{4cm}
\unitlength=1.00mm
\linethickness{0.4pt}
\begin{picture}(141.00,90.00)
\put(30.00,50.00){\line(3,0){20.00}}
\put(50.00,90.00){\line(3,0){20.00}}
\put(100.00,50.00){\line(3,0){20.00}}
\put(120.00,90.00){\line(3,0){20.00}}
\put(30.00,55.00){\makebox(0,0)[cc]{$x$}}
\put(70.00,83.00){\makebox(0,0)[cc]{$y$}}
\put(100.00,55.00){\makebox(0,0)[cc]{$x$}}
\put(140.00,83.00){\makebox(0,0)[cc]{$y$}}
\put(90.00,44.00){\line(5,3){10.00}}
\put(90.00,37.00){\makebox(0,0)[cc]{$x+\mu$}}
\put(50.00,20.00){\makebox(0,0)[cc]{\large{a)}}}
\put(120.00,20.00){\makebox(0,0)[cc]{{\large b)}}}
\put(50.00,50.00){\vector(0,1){20.00}}
\put(50.00,90.00){\line(0,-3){20.00}}
\put(120.00,50.00){\vector(0,1){20.00}}
\put(120.00,70.00){\line(0,3){20.00}}
\put(30.00,49.00){\circle{2.00}}
\put(70.00,89.00){\circle*{2.00}}
\put(52.00,88.00){\vector(0,-1){20.00}}
\put(52.00,48.00){\line(0,3){20.00}}
\put(30.00,48.00){\line(3,0){22.00}}
\put(52.00,88.00){\line(3,0){18.00}}
\put(90.00,43.00){\circle{2.00}}
\put(140.00,89.00){\circle*{2.00}}
\put(122.00,88.00){\line(3,0){18.00}}
\put(122.00,88.00){\vector(0,-1){20.00}}
\put(122.00,48.00){\line(0,3){20.00}}
\put(100.00,48.00){\line(3,0){22.00}}
\put(90.00,42.00){\line(5,3){10.00}}
\end{picture}

\vspace{2cm}

\begin{description}
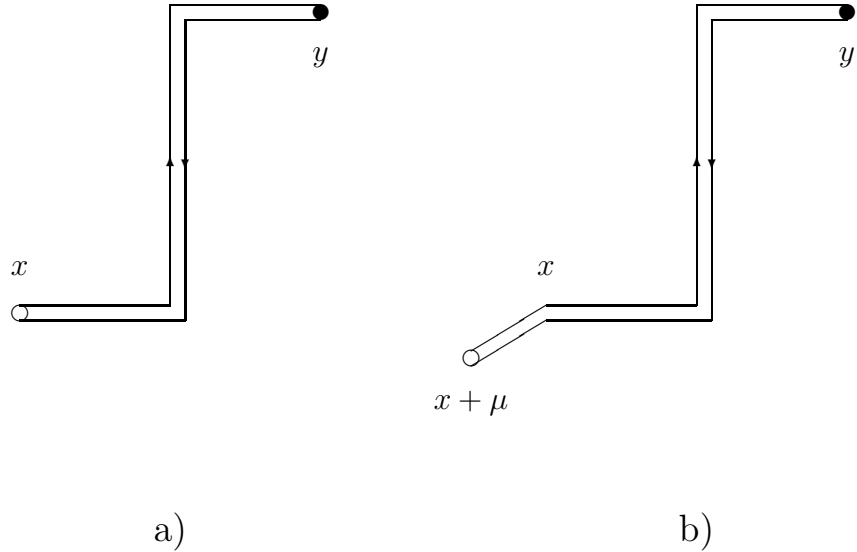
   \item[Fig. 1] \ \ \ The graphic representation for $G_{ij}(C_{xy})$ (a)
   and $G_{ij}(C_{(x+\mu)x}C_{xy})$ (b) entering \eq{sd}. The empty and
   filled circles represent ${\Psi}_i(x)$ (or ${\Psi}_i(x+\mu)$) and
   $\bar{\Psi}_j(y)$, respectively.  The oriented solid lines represent the
   path-ordered products $U(C_{xy})$ and $U(C_{(x+\mu)x}C_{xy})$.  The color
   indices are contracted according to the arrows while the spinor ones
   are associated with circles.  \end{description}

\end{document}